\documentclass[twocolumn,twoside]{article}
\usepackage{graphicx}
\usepackage{epsfig}
\newcommand{\hb}{\\ \hspace*{2ex}}
\newcommand{\hc}{\\ \hspace*{3ex}}
\begin{document}
\title{SIGNATURES OF LARGE-SCALE STRUCTURE OF UNIVERSE IN X-RAYS}
\author{A.V.\,Tugay\\[2mm]
 Taras Shevchenko National University of Kyiv,\hb
 Kyiv, Ukraine, {\em tugay.anatoliy@gmail.com}\\
}
\date{}
\maketitle

ABSTRACT. A new sample of 4299 galaxies with X-ray emission was obtained in this work by cross-correlating 2XMM catalog of X-ray sources with HyperLeda database of galaxies. We analyzed distributions of optical and X-ray fluxes, redshifts and X-ray luminosities for this sample. Virgo and Coma galaxy clusters can be easily detected at redshift-space distribution of X-ray galaxies. X-ray luminosity function of our galaxies shows the evidences of cosmological evolution, even at low redshifts. \\[1mm]
{\bf Key words}: Galaxies: catalogues: X-rays;
databases: XMM-Newton, HyperLeda.\\[2mm]

{\bf 1. Introduction}\\[1mm]

Modern space observatories such as Chandra and XMM-Newton can detect a large number of extragalactic X-ray sources. It is possible now to perform statistical analysis of such sources and study large-scale structure of Universe in X-ray band.
One of the main characteristics of large-scale structure is two-point correlation function for spatial or angular distribution of extragalactic objects. This function was estimated for X-ray active galactic nuclei (AGN) within XMM-LSS project (Elyiv et al., 2012). XMM observations were also used for studying of cosmological evolution of AGN's luminosity function by Ebrero et al. (2009). X-ray galaxy luminosity functions based on Chandra data were obtained by Yencho et al. (2009) and Aird et al. (2010). 

The largest current catalog of X-ray sources is XMM-Newton Serendipitous Source Catalog which is based on the whole XMM-Newton observations archive. Recent version of this catalog, Data Release 3 of 2XMM is available since 2010 (2XMMi-DR3, 2010). This data release contains 262902 automatically detected X-ray sources.
XMM-SSC was previously used for estimating X-ray galaxy luminosity function by Georgakakis \& Nandra (2011). In this work X-ray sources were identified as galaxies by cross-correlation with SDSS sources (Aihara et al., 2011) and checking SDSS spectra. The sample of X-ray galaxies found by this way in redshift interval from 0.1 to 0.8 consist of 209 objects. 

In all previous works samples of X-ray galaxies don't exceed 500 objects. Here we present a new large sample obtained with the largest available now databases of X-ray sources and galaxies: 2XMM and HyperLeda. Now HyperLeda database has 1144990 entries marked as galaxies. Some of them, of course, are misidentifications or false data, the same may be said about some part of 2XMM catalog. Nevertheless, X-ray galaxy sample obtained by cross-correlation of these two large databases may be useful as a pool for statistical studies and for selection of certain interesting objects. In this paper we discuss distribution of initial parameters of our sample and X-ray luminosity - redshift dependence for extragalactic sources. \\[2mm]

{\bf 2. New large sample of X-ray galaxies}\\[1mm]

Since the angular resolution of XMM-Newton is equal to 7 arcsec, we decided to identify each HyperLeda galaxy as X-ray source when there was 2XMM source at distance no more than 7 arcsec from it. This gives us 4299 identifications which we will call "X-ray galaxies". To find the upper limit of the number of X-ray galaxies we enlarged correlation radius to the value of optical radius of HyperLeda galaxy. This procedure increases the number of cross-correlations to 5021. In this work we will consider only the first sample, because the latter should have a lot of misidentifications such as indefinite X-ray source close to center of thin edge-on galaxy that doesn't lie in the disc.
All 4299 galaxies from the first sample has the main X-ray source in the center. In our paper (Tugay\&Vasilenko, 2011) we show that X-ray emission of galactic disc without nuclei is very faint and rarely can be detected. So we can suppose that our new sample consist of AGN's of different types.
Sky distribution of galaxies from our sample in equatorial coordinates is presented at Fig 1. The main density excesses at such sky map are Virgo and Coma galaxy clusters and XMM-LSS region in Cetus (the last is a selection effect caused by intesive XMM-LSS observational program). These regions are marked by circles. Except them and Milky Way zone of avoidance, angular distribution of X-ray galaxies is uniform.
 
 \begin{figure}[h]
 \resizebox{\hsize}{!}{\includegraphics{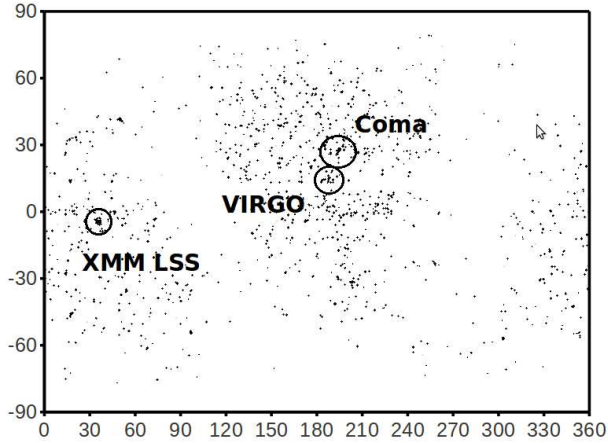}}
 \caption{Sky distribution of X-ray galaxies.}
 \end{figure}

Let's consider spatial redshift-space distribution of our galaxies in Local Supercluster plane. 
This region was observed intensively in Sloan Digital Sky Survey and include the most bright and nearby galaxies, so we should find here the largest density of observed objects. We selected region with $12h<RA<14h$ and $DEC>0$. 242748 galaxies from HyperLeda allows this coordinate constraints. Distribution of these galaxies is shown on Fig. 2. 

 \begin{figure}[h]
 \resizebox{\hsize}{!}{\includegraphics{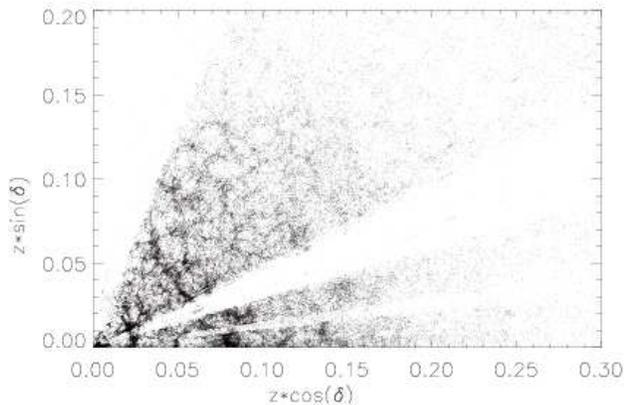}}
 \caption{Redshift-space distribution of optical galaxies in Local Supercluster plane.}
 \end{figure}

The region of supergalactic plane include 778 X-ray galaxies from new sample. At Fig. 3 we can see that only Virgo and Coma clusters can be detected in X-rays in Supergalactic plane. These clusters are the only elements from all large-scale structure that appear in X-rays with modern observational data.

 \begin{figure}[h]
 \resizebox{\hsize}{!}{\includegraphics{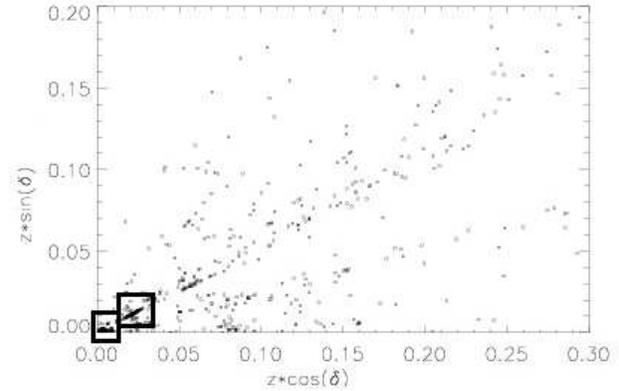}}
 \caption{Redshift-space distribution of X-ray galaxies in Local Supercluster plane. Virgo and Coma Clusters are marked with squares.}
 \end{figure}

Since the details of spatial large-scale structure can not be resolved with our sample, let's pass to luminosity distribution of X-ray galaxies. Fig. 4 presents the distribution of optical and X-ray brightness. u-band magnitudes (most data from SDSS) and X-ray fluxes in whole XMM energy range - 0.2-12 keV are plotted here. According to this distribution our sample seems to be homogeneous.

 \begin{figure}[h]
 \resizebox{\hsize}{!}{\includegraphics{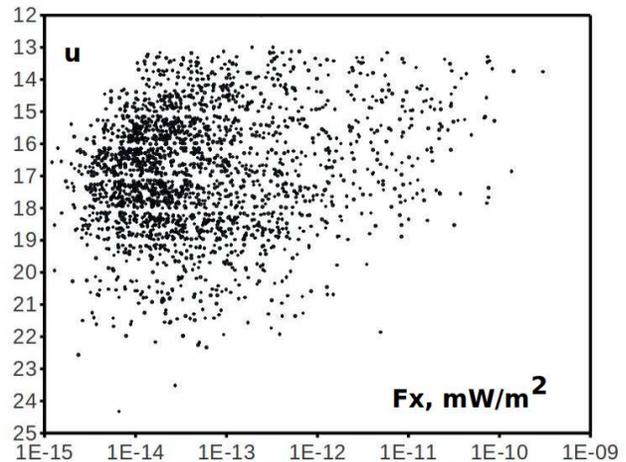}}
 \caption{u-band magnitude against 0.2-12 keV flux for the 2XMM X-ray sources with optical galaxy counterparts.}
 \end{figure}

Distribution of X-ray versus optical ratios and X-ray luminosities is plotted on Fig. 5.  We see that the most galaxies has X-ray luminosity at 0.1\% of optical. Larger ratios and luminosities corresponds to transition from Seyfert galaxies to QSO's and BL Lacs. Note that transition is continuous. Finally, the dependence of X-ray luminosity from redshift is shown in Fig. 6. Lower bound corresponds to the limit of XMM-Newton sensitivity. 

 \begin{figure}[h]
 \resizebox{\hsize}{!}{\includegraphics{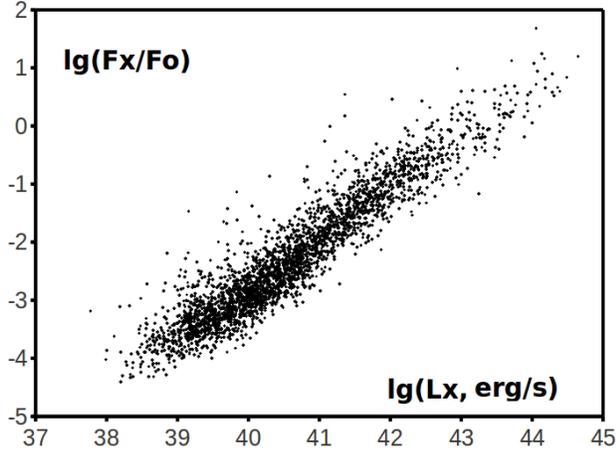}}
 \caption{X-ray to optical flux ratio against X-ray luminosity.}
 \end{figure}

 \begin{figure}[h]
 \resizebox{\hsize}{!}{\includegraphics{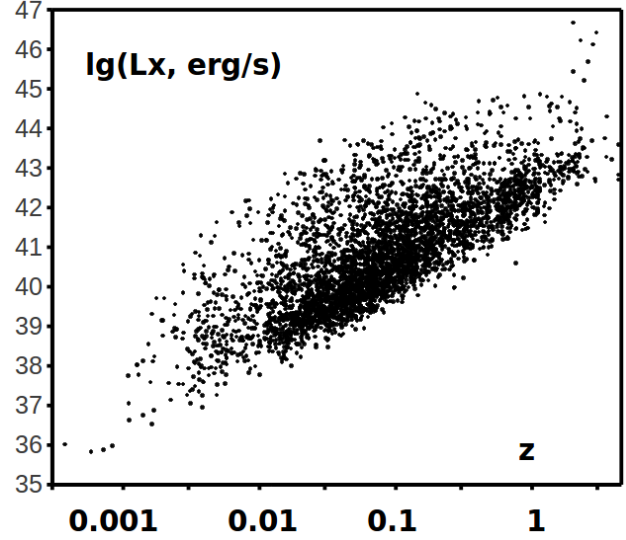}}
 \caption{Redshift versus luminosity plot of our sample. Lower bound of distribution indicates the limit of XMM sensitivity.}
 \end{figure}

To estimate X-ray luminosity function for our sample we calculated the numbers of galaxies in selected intervals of luminosity and redshift distribution. The results are presented in Table 1. Luminosity and redshift intervals corresponds to that from (Ebrero et al, 2009). Note that mean luminosity increases with redshift, that agrees with Ebrero et al. (2009) and other works. This effect is commonly interpreted as cosmological evolution of X-ray AGNs. New sample contain large number of galaxies at lower redshifts, so the evolution effect can be found at Fig. 6 for $z<0.5$.\\[2mm]

\begin{table}[h]
\caption{Luminosity distribution of X-ray galaxies for different redshift ranges.}
\begin{tabular}{crrrr}
\hline
 $lg(L_x, erg/s)$ & 0.01-0.5&0.5-1.0&1.0-2.0&2.0-3.0\\
\hline
41-42 & 770&116& 15& 0\\
42-43 & 365&177& 84&16\\
43-44 & 128& 34& 36&25\\
44-45 &  27& 11& 12& 5\\
\hline
\begin{tabular}{c}
Whole \\ luminosity \\ range \\
\end{tabular}
&3480&340&147&51\\
\hline
\end{tabular}
\end{table}

{\bf 3. Conclusion}\\[1mm]

Although new sample of X-ray galaxies was made using non-homogeneous 2XMM and HyperLeda databases, its object shows uniform distribution both in space and luminocities. XMM-Newton observations can not reveal cell-like 
\linebreak\vfill\pagebreak\noindent
large-scale structure of Universe but are useful for investigation X-ray luminosity functions and individual X-ray sources. \\[2mm]

\indent
{\bf References\\[2mm]}
Aihara H., Allende Prieto C., An D. et al.: 2011, {\it Ap. J. S.}, {\bf 193}, 29.\\
Aird J., Nandra K., Laird E.S.: 2010, {\it MNRAS}, {\bf 401}, 2531.\\
Ebrero J., Carrera F.J., Page M.J. et al.: 2009, {\it A\& A.}, {\bf 493}, 55.\\ 
Elyiv A., Clerc N., Plionis M. et al.: 2012, {\it A\& A.}, {\bf 537}, 131.\\
Georgakakis A., Nandra K.: 2011, {\it MNRAS}, {\bf 414}, 992.\\
Tugay A.V., Vasilenko A.A.: 2011, {\it Odessa Astronomicel Publications}, {\bf 24}, 72.\\
Yencho B., Barger A.J., Trouille L. et al.: 2009, {\it Ap. J.}, {\bf 698}, 380.\\
2XMMi-DR3 (XMM-SSC): 2010,\hc {\it http://vizier.u-strasbg.fr/viz-bin/VizieR?-source=IX\%2F41}

HyperLeda, {\it http://leda.univ-lyon1.fr}
\end{document}